\documentclass{ws-procs9x6}
\usepackage{psfrag}
\def\citebk#1{$\mbox{[\raisebox{-1.85mm}[0mm][0mm]
  {\Large\cite{#1}}\hspace{-0.1mm}]}$}

\def\citebkcap#1{[\raisebox{-1.5mm}[0mm][0mm]
  {\large\cite{#1}}\hspace{-0.1mm}]}

\def\citeno#1{\raisebox{-2mm}[0mm][0mm]
  {\Large\cite{#1}}\hspace{-0.1mm}}

\newcommand{\be}{\begin{equation}}
\newcommand{\ee}{\end{equation}}
\newcommand{\bdm}{\begin{displaymath}}
\newcommand{\edm}{\end{displaymath}}
\newcommand{\bea}{\begin{eqnarray}}
\newcommand{\eea}{\end{eqnarray}}
\newcommand{\no}{\nonumber \\}
\newcommand{\fs}{\; \; .}
\newcommand{\co}{\; \; ,}

\newcommand{\lvac}{\langle 0|\,}
\newcommand{\rvac}{\,|0\rangle}

\newcommand{\Fem}{F}
\newcommand{\s}{s_2}
\begin{document}

\title{Electromagnetic form factor of the pion}

\author{H.~Leutwyler}

\address{Institute for
Theoretical Physics, University of Bern,\\ Sidlerstr. 5, CH-3012 Bern,
Switzerland\\
E-mail: leutwyler@itp.unibe.ch}

\maketitle

\abstracts{The Standard Model prediction for the magnetic moment of the muon
  requires a determination of the electromagnetic form factor of the pion 
  at high precision. It is shown that the recent progress in 
  $\pi\pi$ scattering allows us to obtain an accurate representation of this
  form factor on the basis of the data on 
  $e^+e^-\rightarrow\pi^+\pi^-$. The same method also applies to the form
  factor of the weak vector current, where the data on the decay
  $\tau\rightarrow\pi^-\pi^0\,\nu_\tau$ are relevant. Unfortunately, however,
  the known sources of isospin breaking do not explain the difference
  between the two results. The discrepancy implies that the Standard Model
  prediction for the magnetic moment of the muon is currently subject to a
  large uncertainty. \begin{center}Talk given at the Workshop {\it Continuous
  Advances in QCD 2002/Arkadyfest}\\ in honor of the 60th birthday of Arkady
  Vainshtein, Minneapolis, May 2002.\end{center}}

\section{Motivation: magnetic moment of the muon}
The fabulous precision reached in the measurement of the muon magnetic 
moment \citebk{experiment} allows a thorough test of the Standard Model.
The prediction that follows from the Dirac equation, $\mu
=e\,\hbar/2 m_\mu$, only holds to leading
order in the expansion in powers 
of the fine structure constant $\alpha$. It is customary to write the
correction in the form
\be\mu =\frac{e\,\hbar}{2\,m_\mu}\,(1+a)\fs\ee
Schwinger was able to calculate the
term of first order in $\alpha$, which stems from the triangle graph in 
fig.~1a
and is universal \citebk{Schwinger},
\be\label{Schwinger} a=\frac{\alpha}{2\pi}+O(\alpha^2)\fs\ee
The contributions of $O(\alpha^2)$ can also unambiguously be calculated,
except for the one from hadronic vacuum polarization, indicated by the graph
in fig.~1e. It is analogous to the contributions generated by leptonic
vacuum  polarization in figs.~1b, 1c and 1d,
\begin{figure}
\begin{tabular}{ccccc}
\includegraphics[width=2cm]{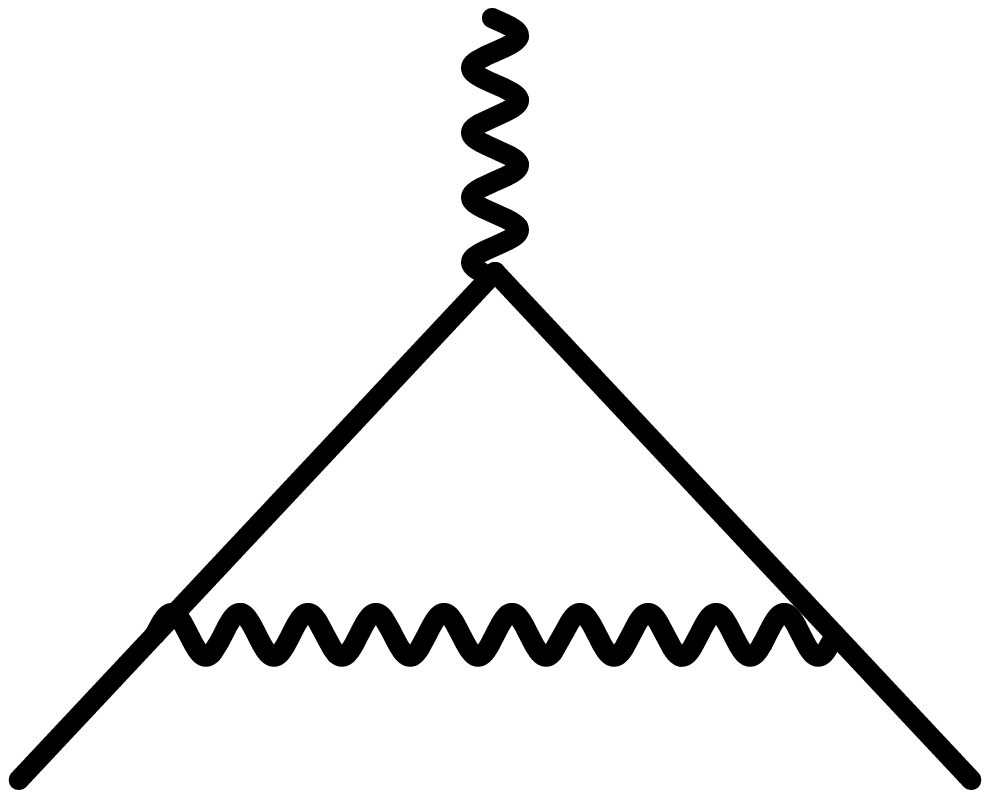}&
{\psfrag{e}{\raisebox{0.1em}{$\!e$}}
\includegraphics[width=1.9cm]{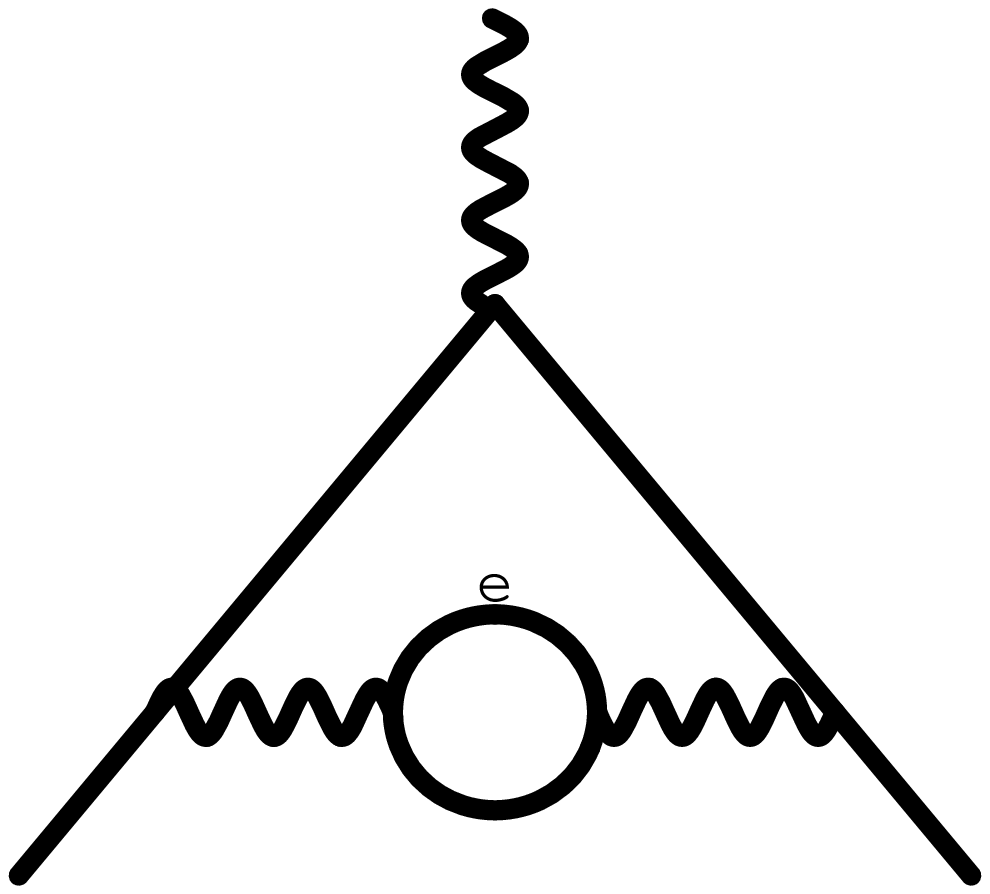}}&
{\psfrag{e}{\raisebox{0.1em}{$\!\mu$}}
\includegraphics[width=1.9cm]{graph3.eps}}&
{\psfrag{e}{\raisebox{0.1em}{$\!\tau$}}
\includegraphics[width=1.9cm]{graph3.eps}}&
\includegraphics[width=1.9cm]{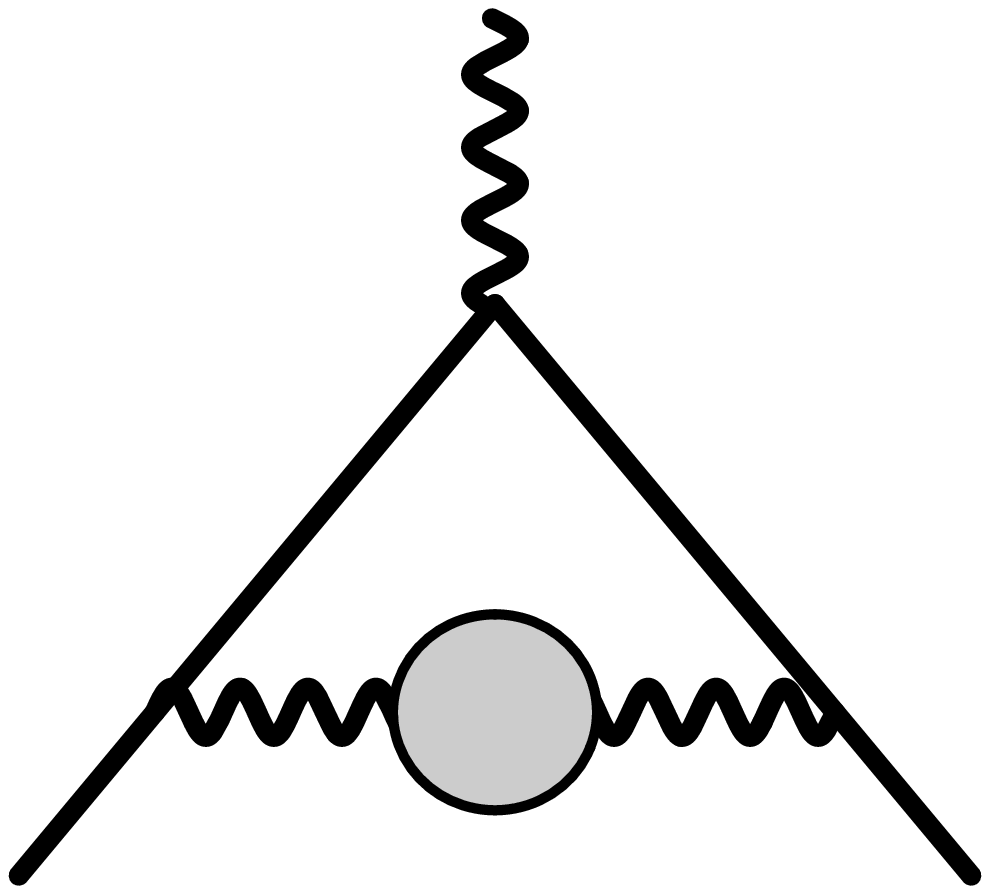}\\
1a& 1b&1c&1d&1e
\end{tabular}
\end{figure}
but involves
quarks and gluons instead of leptons. All of these graphs may be viewed as
arising from vacuum polarization in the photon propagator:
\bdm D_{\mu\nu}(q)=\frac{g_{\mu\nu}\,Z}{q^2\{1+\Pi(q^2)\}}+\mbox{gauge
  terms}\fs 
\edm
The expansion of the self energy function $\Pi(t)$ in powers of $\alpha$
starts with 
\bdm \Pi(t)=\alpha\,\Pi^{(0)}(t)+\alpha^2\,\Pi^{(1)}(t)+\ldots\edm
It is normalized by $\Pi(0)=0$.
The leading term can be pictured as

\begin{figure}
$\Pi^{(0)}(t)=$\rule{0.3em}{0em}
\raisebox{-0.2em}{\psfrag{e}{\raisebox{0.1em}{$\!e$}} 
\includegraphics[width=0.7cm]{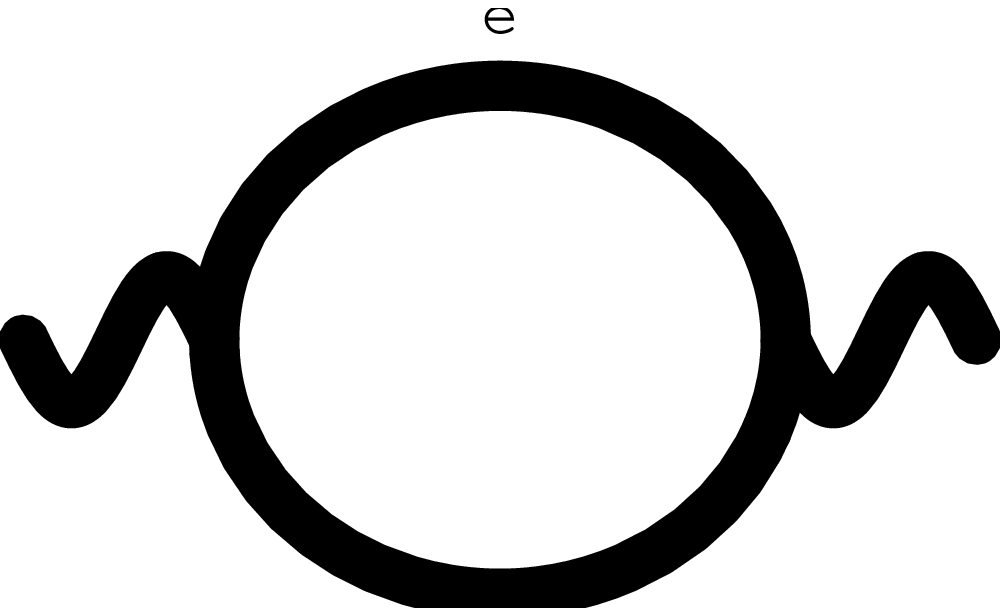}} \rule{0.3em}{0em}+
\rule{0.3em}{0em}\raisebox{-0.2em}
{\psfrag{e}{\raisebox{0.1em}{$\!\mu$}}
\includegraphics[width=0.7cm]{graph4.eps}}\rule{0.3em}{0em} +
\rule{0.3em}{0em}\raisebox{-0.2em}
{\psfrag{e}{\raisebox{0.1em}{$\!\tau$}}
\includegraphics[width=0.7cm]{graph4.eps}}\rule{0.3em}{0em}+
\rule{-0.1em}{0em}\raisebox{-0.3em}{
\includegraphics[width=0.7cm]{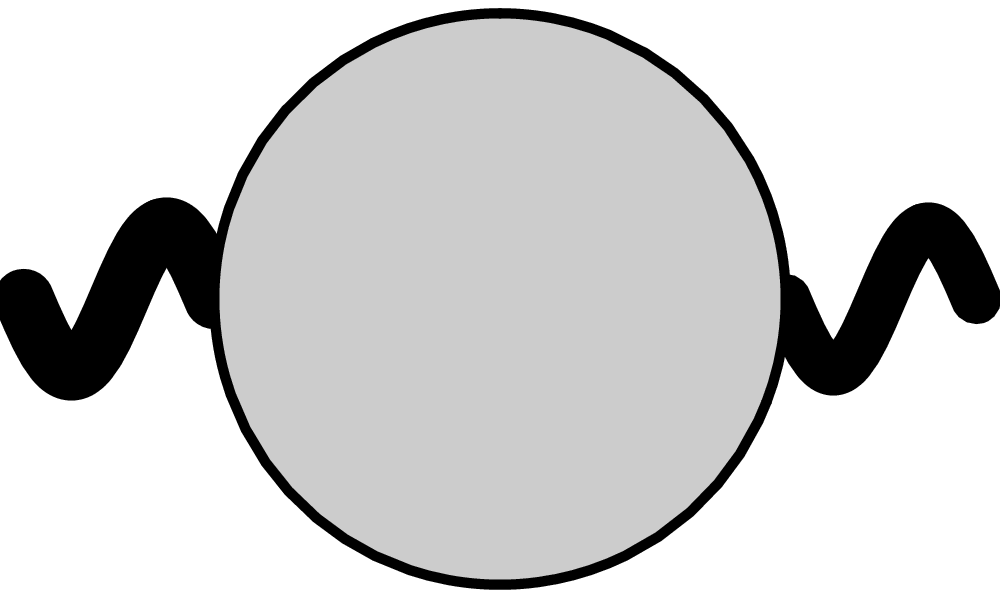}}
\end{figure}

\vspace{-1em}
\noindent As shown in ref.~\citebk{Calmet}, the modification of the Schwinger
formula (\ref{Schwinger}) that is generated by vacuum polarization can be
represented in compact form:\footnote{
The formula only makes sense
in the framework of the perturbative expansion \citebkcap{Lautrup}.
The contribution generated by an electron loop, for instance, grows
logarithmically at large momenta and tends to $-\infty$ in the spacelike
region, 
\bdm\Pi^{(0)}_e(t)= -\frac{1}{3\pi}\ln
\frac{(-t)}{M_1^2}+O\left(\frac{M_e^2}{t}\right)
\co\hspace{2em}M_1=e^{\frac{5}{6}}\,M_e\fs\edm
Hence $1+\alpha \,\Pi^{(0)}_e(t)$ contains a zero in the vicinity of
$t=-\exp (3\pi/\alpha)\,M_1^2\simeq-10^{555}\,\mbox{GeV}^2$ (graphs 1c, 1d and
1e push the zero towards slightly smaller values).
At academically high energies, the photon propagator thus develops a
''Landau pole'', reflecting the fact that the U(1) factor of the Standard
Model does not give rise to an asymptotically free gauge theory. In the present
context, however, 
this phenomenon is not relevant -- we are concerned with the low energy
structure of the Standard Model.} 
\be\label{avacpol} a=\frac{\alpha}{\pi}\int_0^1\!\!\!
dx\,\frac{1-x}{1+\Pi(t_x)}\co 
\hspace{2em}
t_x=-\frac{x^2m_\mu^2}{1-x}\fs\ee
Expanding this formula in powers of $\alpha$, we obtain
\bdm a=\frac{\alpha}{\pi}\int_0^1\!\!dx\,(1-x)-\frac{\alpha^2}{\pi}
\int_0^1\!\!\! dx\,(1-x)\,\Pi^{(0)}(t_x)+O(\alpha^3)\edm
The first term indeed reproduces the  Schwinger formula
(\ref{Schwinger}), which corresponds to graph 1a. The term linear in
$\Pi^{(0)}(t)$ accounts for graphs 1b to 1e. The contribution involving the square
of $\Pi^{(0)}(t)$ describes the one-particle reducible graphs with two 
bubbles, etc. 

The vacuum polarization due to a lepton loop is given by $(\ell=e,\mu,\tau)$
\bea\label{Rell} \Pi_\ell^{(0)}(t)&=&\frac{t}{3\pi}\int_{4 M_\ell^2}^\infty\!\!
ds\frac{R_\ell(s)}{s(s-t)}\co\\ 
R_\ell(s)&=& \sqrt{1-\frac{4M_\ell^2}{s}}\left(1+\frac{2M_\ell^2}{s}\right)\fs
\nonumber\eea 
The hadronic contribution cannot be calculated
analytically, but it can be 
expressed in terms of the cross section of the reaction
$e^+e^-\rightarrow\mbox{hadrons}$. More precisely, the leading term in the
expansion of this cross section in powers of $\alpha$,
\bdm\sigma_{e^+e^-\rightarrow h}=\alpha^2 \sigma_{e^+e^-\rightarrow
  h}^{(0)}+O(\alpha^3)\co\edm   
is relevant. In terms of this quantity the expression reads
\bea \Pi_h^{(0)}(t)&=&\frac{t}{3\pi}\int_{4 M_\pi^2}^\infty\!\!
ds\frac{R_h(s)}{s(s-t)}\co\\
R_h(s)&=& \frac{3 s}{4\pi}\,\sigma_{e^+e^-\rightarrow h}^{(0)}(s)\fs
\nonumber\eea
At low energies, where the final state necessarily consists of two pions, the
cross section is given by the square of the electromagnetic form factor of the
pion,
\be\label{Rh}
R_h(s)=\frac{1}{4}\left(1-\frac{4M_\pi^2}{s}\right)^{\frac{3}{2}}\, 
|\Fem(s)|^2\co\hspace{2em} s<9\,M_\pi^2\fs\ee 

Numerically, the contribution from hadronic vacuum polarization to the
magnetic moment of the muon amounts to $a_{\mbox{\tiny hvp}}\simeq 700\times
10^{-10}$.  This is a small fraction of the total,
 $ a=11\,659\,203\, (8) \times 10^{-10}$ \citebk{experiment},
but large compared to the experimental uncertainty: a 
determination of $a_{\mbox{\tiny hvp}}$ to about 1\% is required 
for the precision
of the Standard Model prediction to match the experimental one. Since
the contribution from hadronic vacuum polarization is dominated by the one
from the two pion states, this means that the pion form factor is needed to an
accuracy of about half a percent. 

\newpage
\section{Comparison of leptonic and hadronic contributions} 
Graphically, the formula (\ref{Rh}) amounts to
\begin{figure}
\hspace{3em}\includegraphics[width=2cm]{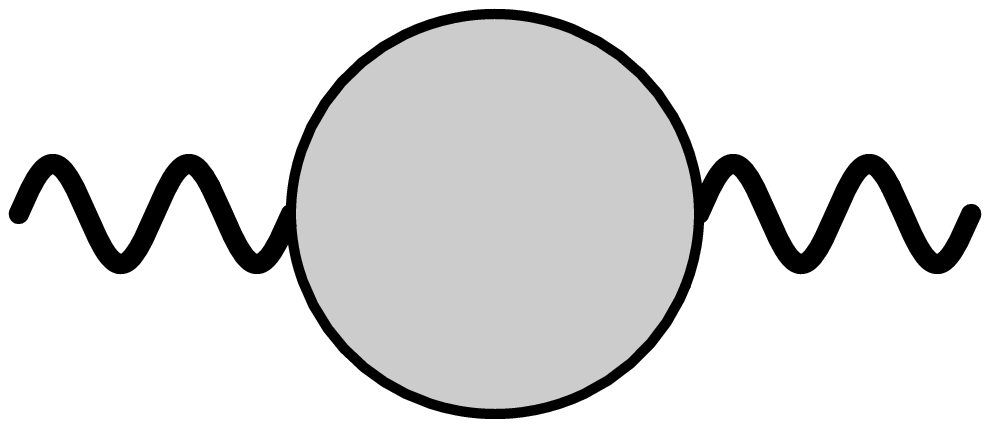}\hspace{1em} 
\raisebox{0.8em}{$\Rightarrow$}\hspace{1em} 
{\psfrag{pi}{\raisebox{0.1em}{$\!\pi$}}
\includegraphics[width=2cm]{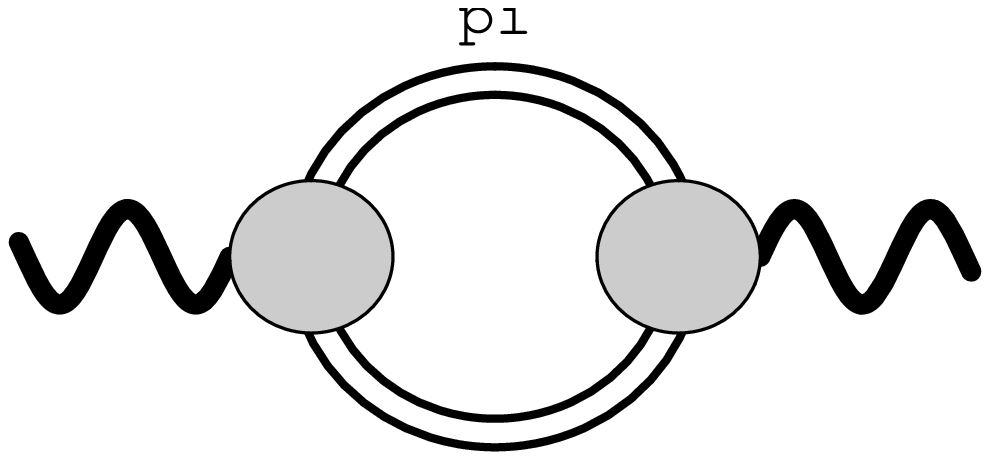}}
\end{figure}

\noindent There are three differences between the pionic loop integral and
those belonging to the lepton loops:
\begin{itemize}
\item the masses are different
\item the spins are different
\item the pion is composite -- the Standard Model leptons are
  elementary 
\end{itemize}
The compositeness manifests itself in the occurrence of the form factor
$\Fem(s)$, which generates an enhancement: at the $\rho$ peak, $|\Fem(s)|^2$ 
reaches
values of order 45. The remaining difference in the expressions for the
quantities $R_\ell(s)$ and $R_h(s)$ in eqs.~(\ref{Rell}) and (\ref{Rh})
originates in the fact that the leptons carry spin $\frac{1}{2}$, while the
spin of the pion vanishes. Near threshold, the angular momentum barrier
suppresses the function $R_h(s)$ by three powers of momentum, while
$R_\ell(s)$ is proportional to the first power. The suppression largely
compensates the enhancement by the form factor -- by far the most important
property is the mass: in units of $10^{-10}$, the contributions due to the
$e,\mu$ and $\tau$ loops are 59040.6, 846.4 and 4.2, respectively, to be
compared with the  
700 units from hadronic vacuum polarization. The latter is comparable to the
one from the muon -- in accordance with the fact that
the masses of pion and muon are similar. 

\section{Pion form factor}\label{pion form factor}
In the following, I disregard the electromagnetic interaction --
the discussion concerns the properties of the form
factor in QCD. I draw from ongoing work
carried out in collaboration with Irinel Caprini, Gilberto Colangelo, Simon
Eidelman, J\"urg Gasser and Fred Jegerlehner. 

The systematic low energy analysis of the form factor 
based on chiral perturbation theory \citebk{GL form factor} has been worked
out to two loops \citebk{two loops}. This approach,
however, only covers the threshold region. The range of
validity of the representation can be extended to higher energies by means of
dispersive methods \citebk{dispersive methods}, which exploit the constraints
imposed by analyticity and unitarity. 
Our approach is very similar to the one of de Troc\'oniz 
and Yndurain $\mbox{\citebk{Troconiz Yndurain}}$. For a thorough discussion of
the mathematical framework, I refer to Heyn and Lang \citebk{Heyn Lang}.

We represent the form factor as a
product of three functions that account for the prominent singularities
in the low energy region:
\be\label{three factors} \Fem(s)=G_2(s)\times G_3(s)\times G_4(s)\fs\ee
The index has to do with the number of pions that generate the relevant
discontinuity: two in the case of $G_2$,  
three for $G_3$ and four or more for $G_4$.

The first term represents the familiar
Omn\`es factor that describes the branch cut 
due to $\pi^+\pi^-$ intermediate states (states with two
neutral pions do not contribute, because the matrix
element $\langle\pi^0\pi^0\,\mbox{out}\,|\,j^\mu\rvac$ vanishes, on account of
Bose 
statistics). The corresponding branch point singularity
is of the type $\mbox{Im}\,G_2(s)\sim (s-s_2)^{\frac{3}{2}}$, with $s_2 =
4M_\pi^2$.  The Watson final state interaction theorem implies that,
in the elastic region, $4M_\pi^2<s<9M_\pi^2$, the phase of the form factor
is given by the P-wave phase shift
of the elastic scattering process $\pi^+\pi^-\rightarrow\pi^+\pi^-$. Denoting
this phase shift by $\delta(s)$, the explicit expression for the Omn\`es
factor reads:
\be\label{G2} G_2(s)=\exp \left\{\frac{s}{\pi}\int_{4M_\pi^2}^\infty
\frac{dx\;\delta(x)}{x\,(x-s)}\right\} \fs\ee

The function $G_3(s)$ contains the singularities generated by $3\pi$
intermediate states:  $G_3(s)$ is analytic except for a cut starting at
$s_3=9M_\pi^2$, with a branch point singularity of the type
$\mbox{Im}\,G_3(s)\sim (s-s_3)^4$.
If isospin symmetry were exact, the form factor would not
contain such singularities:
in the limit $m_u=m_d$, the term $G_3(s)$ is equal to 1.
Indeed, isospin is nearly
conserved, but the occurrence 
of a narrow resonance with the proper quantum numbers strongly enhances the
effects generated by isospin breaking: the form factor contains a pole close
to the real axis,
\be\label{G3}
G_3(s)=1+\epsilon\,\frac{s}{s_\omega-s}+\ldots\hspace{2em}
s_\omega=(M_\omega-\mbox{$\frac{1}{2}$}\,i\,\Gamma_\omega)^2\fs\ee
This implies that, in the vicinity of $s=M_\omega^2$, the form factor rapidly
varies, both in magnitude and in phase. The pole term cannot stand by itself
because it fails to be real in the 
spacelike region. We replace it by a dispersion integral with the proper
behaviour at threshold, but this is inessential: in the experimental range,
the representation for
$G_3(s)$ that we are using can barely be distinguished from the
pole approximation (\ref{G3}). 

Isospin breaking also affects the scattering amplitude. In particular, it
gives rise to the inelastic reaction
$2\pi\rightarrow 3\pi$, with an amplitude proportional to $m_u-m_d$. Hence
unitarity implies that, in the region $9M_\pi^2<s<16M_\pi^2$, 
the elasticities of the partial waves are less than 1. Numerically, the effect
is tiny, however, because it is of second order in $m_u-m_d$. To a very high
degree of accuracy, the first two terms in eq.~(\ref{three factors})
thus account for all singularities below $s_4 = 16 M_\pi^2$ -- the function
$G_4(s)$ is analytic in the plane cut from $s_4$ to $\infty$. Phase space
strongly suppresses the strength of 
the corresponding branch point singularity: 
$\mbox{Im}\,G_4(s)\propto (s-s_4)^\frac{9}{2}$. A significant
discontinuity due to inelastic channels only 
manifests itself for $s>s_{in}=(M_\omega+M_\pi)^2$.  

We analyze the background term $G_4(s)$
by means of a conformal mapping. The transformation
\be\label{conformal map}
z=\frac{\sqrt{s_{in}-s_1}\,-\sqrt{s_{in}-s}}{\sqrt{s_{in}-s_1}\,+
\sqrt{s_{in}-s}}\ee  
maps the $s$-plane cut along $s>s_{in}$ onto the unit disk in the
$z$-plane, so that the Taylor series expansion in powers of $z$ converges
on the entire physical sheet, irrespective of the value of the arbitrary
parameter $s_1$. We truncate this series after the first few
terms, thus approximating the function $G_4(s)$ by a low order polynomial in
$z$.  

\section{Roy equations}
The crucial element in the above representation is the phase $\delta(s)$.  
The main difference between our analysis and the one in 
ref.~\citebk{Troconiz Yndurain} concerns the input used to describe the
behaviour of this phase. In fact, during the last two or three years,
our understanding of the $\pi\pi$ scattering amplitude has made a quantum
jump. As a result of theoretical work [\citeno{ACGL}--\citeno{Descotes}], the low energy behaviour of the S- and
P-waves is now known 
to an amazing accuracy -- to my knowledge,
$\pi\pi$ scattering is the only field in strong
interaction physics where theory is ahead of experiment. 

The method used to implement the requirements of analyticity, unitarity and
crossing symmetry is by no means new. As shown by Roy more than
30 years ago  \citebk{Roy}, these properties of the scattering amplitude
subject the partial waves to a set of coupled integral equations.
These equations involve two subtraction constants, which may be
identified with the two $S$--wave scattering lengths $a_0^0$, $a_0^2$. 
If these two constants are given, the Roy equations allow us to calculate the
scattering 
amplitude in terms of the imaginary parts above the "matching point"
$E_m=0.8\,\mbox{GeV}$. The available
experimental information suffices to evaluate the relevant dispersion
integrals, to within small uncertainties \citebk{ACGL,Descotes}. In this
sense,  $a_0^0$, $a_0^2$ represent the essential parameters in low energy
$\pi\pi$ scattering. 

As will be discussed in some detail in the next section, chiral symmetry
predicts the values of the two subtraction constants 
and thereby turns the Roy equations into a framework that 
fully determines the low energy behaviour of the $\pi\pi$ scattering amplitude.
In particular, the P-wave scattering length and effective range
are predicted very accurately:
$a_1^1=0.0379(5) \,M_\pi^{-2}$ and $b_1^1=0.00567(13)\,M_\pi^{-4}$. The manner
in which the P-wave phase shift  
passes through 90$^\circ$ when the energy reaches the mass of the $\rho$
is specified within the same framework, as well as the behaviour of the
two S-waves.  The analysis reveals, for instance, that the isoscalar S-wave
contains a pole on the second sheet and the position can be
calculated rather accurately: the pole occurs at
$E=M_\sigma-\frac{1}{2}\,i\,\Gamma_\sigma$,  with
$M_\sigma=470\pm 30\;\mbox{MeV}$, $\Gamma_\sigma=590\pm 40\;\mbox{MeV}$
\citebk{CGL}, etc.  

Many papers based on alternative approaches can be found in the 
literature.  Pad\'e approximants, for instance, continue to enjoy popularity
and the ancient idea that the $\sigma$ pole  
represents the main feature in the isoscalar S-wave also found new adherents
recently. Crude models such as these may be of interest in connection with
other processes where the physics yet remains to be understood, but 
for the analysis of the $\pi\pi$ scattering amplitude, they cannot compete
with the systematic approach based on analyticity and chiral symmetry.
In view of the precision required in the determination of the pion form
factor, ad hoc models are of little use, because the theoretical uncertainties
associated with these are too large.

\section{Prediction for the $\pi\pi$ scattering lengths}
Goldstone bosons of zero momentum do not interact:  if the
quark masses $m_u,m_d$ are turned off, the S-wave scattering lengths
disappear, $a_0^0,a_0^2\rightarrow 0$. Like the mass of the pion, 
these quantities represent effects that arise from the breaking of the chiral
symmetry generated by the quark masses. In fact, as shown by Weinberg
\citebk{Weinberg 1966}, $a_0^0$ and $a_0^2$ are proportional to the square of
the pion mass
\bdm a_0^0=\frac{7M_\pi^2}{32 \pi F_\pi^2}+O(M_\pi^4)\co\hspace{2em} 
a_0^2=-\frac{M_\pi^2}{16 \pi F_\pi^2}+O(M_\pi^4)\fs\edm
The corrections of order $M_\pi^4$ contain chiral logarithms. In the case of
$a_0^0$, the logarithm has an unusually large coefficient
\bdm a_0^0=\frac{7M_\pi^2}{32 \pi
  F_\pi^2}\left\{1+\frac{9}{2}\,\frac{M_\pi^2}{(4\pi
    F_\pi)^2}\,\ln\frac{\Lambda_0^2}{M_\pi^2} +O(M_\pi^4)\right\}\fs\edm
This is related to the fact that in the channel with $I=0$, chiral symmetry
predicts a strong, attractive, final state interaction. The scale $\Lambda_0$
is determined by the coupling constants of the effective
Lagrangian of $O(p^4)$:
\bdm \frac{9}{2}\ln\frac{\Lambda_0^2}{M_\pi^2}=\frac{20}{21}\,
 \mbox{\large$\bar{\ell}_1$}+\frac{40}{21}\,
\mbox{\large$\bar{\ell}_2$}-\frac{5}{14}\,
\mbox{\large$\bar{\ell}_3$}+
2\,\mbox{\large$\bar{\ell}_4$}+\frac{5}{2}\fs\edm 
The same coupling constants also determine the first order correction in the
low energy theorem for $a_0^2$.  

The couplings $\bar{\ell}_1$ and $\bar{\ell}_2$ control the momentum
dependence of the scattering amplitude at first nonleading order. Using the
Roy equations, these constants can be determined very accurately \citebk{CGL}.
The terms $\bar{\ell}_3$ and $\bar{\ell}_4$, on the other hand, 
describe the dependence of the scattering amplitude on the quark masses --
since these cannot be varied experimentally,  $\bar{\ell}_3$ and
$\bar{\ell}_4$ cannot be
determined on the basis of $\pi\pi$ phenomenology. 
The constant $\bar{\ell}_3$ 
specifies the correction in the Gell-Mann-Oakes-Renner relation  \citebk{GMOR},
\be\label{Mpi} M_\pi^2= M^2\left\{1-\frac{1}{2}\,\frac{M^2}{(4\,\pi
      F)^2}\;\bar{\ell}_3+O(M^4)\right\}\fs\ee
Here $M^2$ stands for the term linear in the quark masses,
\be\label{GMOR} M^2=(m_u+m_d)\,|\lvac \bar{u} u\rvac|\,\frac{1}{F^2}\ee
($F$ and $\lvac \bar{u}u\rvac$  are the values of the pion decay constant and
the quark condensate in the chiral limit, respectively). 
The coupling constant $\bar{\ell}_4$ occurs in the analogous expansion for 
$F_\pi$, 
\be\label{Fpi}
F_\pi =  F\left\{1+\frac{M^2}{(4\,\pi F)^2}\;\bar{\ell}_4
+O(M^4)\right\}\fs\ee
A low energy theorem relates it to the scalar radius
of the pion \citebk{GL},
\be\label{radius} 
\langle r^2\rangle\!\rule[-0.3em]{0em}{0em}_s=\frac{6}{(4\pi F)^2}\left\{
\bar{\ell}_4-\frac{13}{12}+O(M^2)\right\}\fs\ee
The dispersive analysis of the scalar pion form factor in
ref.~\citebk{CGL} leads to
\be\label{nradius} \langle r^2\rangle\!\rule[-0.3em]{0em}{0em}_s=0.61\pm0.04\;\mbox{fm}^2
\fs\ee
The constants $\bar{\ell}_1,\ldots \bar{\ell}_4$ depend logarithmically on the
quark masses:
\bdm \bar{\ell}_i=\ln\frac{\Lambda_i^2}{M^2}\co\hspace{2em}i=1,\ldots,4\edm
In this notation, the above value of the scalar radius amounts to 
\be\label{Lambda4} \Lambda_4 =1.26 \pm 0.14 \;\mbox{GeV}\fs\ee
Unfortunately, the constant $\bar{\ell}_3\leftrightarrow\Lambda_3$
is not known with comparable precision. The crude estimate for
$\bar{\ell}_3$ given in ref.~\citebk{GL} corresponds to
\be\label{Lambda3} 0.2\;\mbox{GeV}<\Lambda_3< 2\;\mbox{GeV}\fs\ee
\setcounter{figure}{1}
\begin{figure}[thb]
\psfrag{a00}{\raisebox{-1em}{\Large$a_0^0$}}
\psfrag{a20}{\Large$a_0^2$}

\vspace{-3em}
\hspace{-1em}\includegraphics[width=9cm,angle=-90]{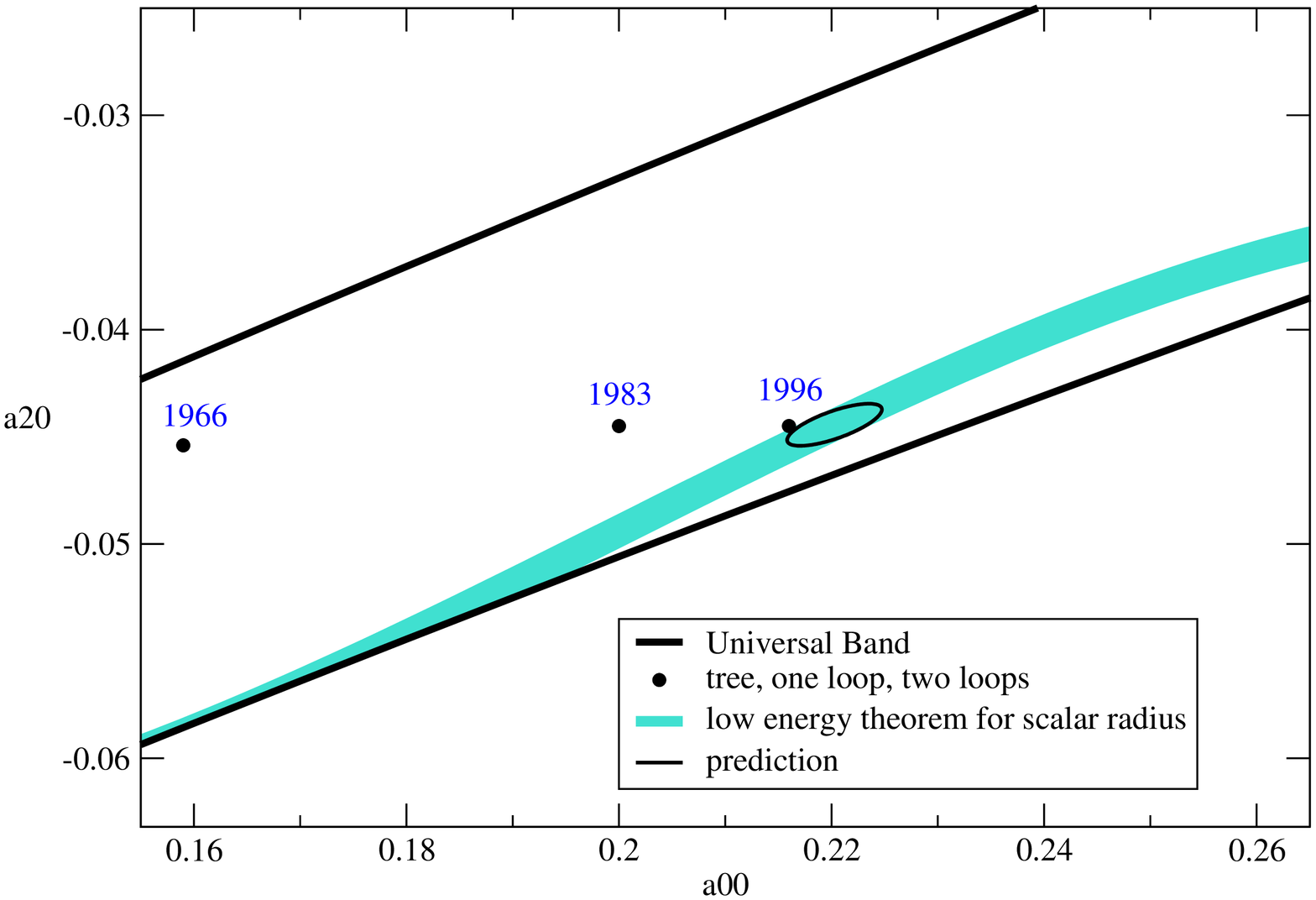}

\caption{Prediction for the S-wave $\pi\pi$ scattering lengths}
\end{figure}
It turns out, however, that the contributions from $\bar{\ell}_3$ are very
small, so that 
the uncertainty in $\Lambda_3$ does not strongly affect the
predictions for the scattering lengths. This is shown in fig.~2, where 
the values of $a_0^0$, $a_0^2$ predicted by ChPT are indicated as a small
ellipse. 

\section{Experimental test}\label{experimental test}
Stern and collaborators \citebk{KMSF}
pointed out that "Standard" ChPT relies on a hypothesis that calls for 
experimental test.
Such a test has now been performed and I wish to briefly describe this
development. 

The hypothesis in question is the assumption that the quark
condensate represents the leading order parameter of the spontaneously broken 
chiral symmetry. More specifically, the standard analysis assumes that
the term linear in the quark masses dominates the expansion of
$M_\pi^2$. According to the
Gell-Mann-Oakes-Renner relation (\ref{GMOR}), this term is proportional to the
quark condensate, which 
in QCD represents the order parameter of lowest dimension. The dynamics of the
ground state is not well understood. The question raised by Stern et al.~is
whether, for one reason or the other,
the quark condensate might turn out to be small, so that the 
Gell-Mann-Oakes-Renner formula would fail -- the "correction" might be 
comparable to or even larger
than the algebraically leading term. 

According to eq.~(\ref{Mpi}), the
correction is determined by the effective coupling constant $\bar{\ell}_3$.
The estimate (\ref{Lambda3}) implies that the correction amounts to at most
4\% of the leading term, but this does not answer the question, because that
estimate is based on the standard framework, where 
$\lvac\bar{u}u\rvac$ is assumed to represent the leading order parameter. 
If that estimate is discarded and $\bar{\ell}_3$ is treated as a free
parameter ("Generalized" ChPT), the scattering lengths cannot be predicted 
individually, but the low energy theorem (\ref{radius})
implies that -- up to corrections of next-to-next-to leading order -- 
the combination $2a_0^0-5a_0^2$ is determined by the scalar
radius:
\bdm 2 a_0^0-5a_0^2=\frac{3M_\pi^2}{4\,\pi F_\pi^2}\left\{1+
\frac{M_\pi^2\langle r^2\rangle\!\rule[-0.3em]{0em}{0em}_s}{3}+\frac{41
  M_\pi^2}{192\,\pi^2F_\pi^2}+O(M_\pi^4)\right\}\fs\edm
The resulting correlation between $a_0^0$ and $a_0^2$ 
is shown as a narrow strip in fig.~2 (the strip
is slightly curved because the figure accounts for the corrections of
next-to-next-to leading order).  

In view of the correlation between $a_0^0$ and $a_0^2$, the data taken by the
E865-collaboration at Brookhaven \citebk{Brookhaven} allow a significant 
test of the Gell-Mann-Oakes-Renner relation. The final state interaction
theorem implies that the phase of the form factors relevant for the
decay $K^+\rightarrow \pi^+\pi^- e^+\bar{\nu}_e$ is determined by the elastic
$\pi\pi$ scattering amplitude. Conversely, the phase difference
$\delta_0^0-\delta_1^1$ can be measured in this decay. The analysis of the
$4\cdot 10^5$ events of this type collected by E865 leads to the 
round data points
in fig.~3, taken from ref.~\citebk{PRLCGL} (the triangles represent the
$K_{e_4}$ data collected in the seventies). 
\begin{figure}[tb]
\begin{center}
\includegraphics[width=9 cm]{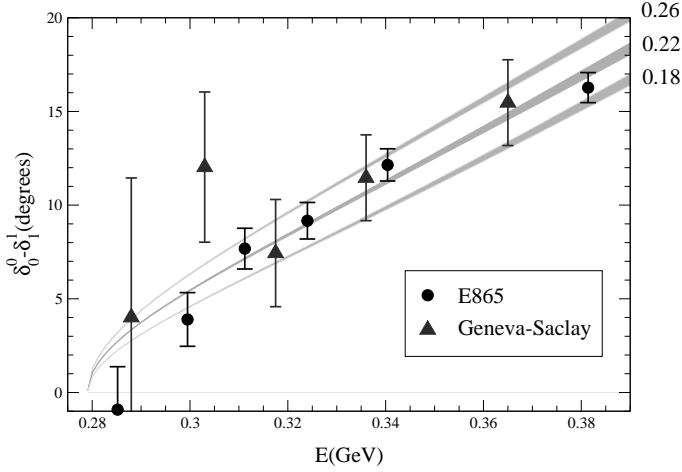}
\end{center}
\caption{Interpretation of the data on the phase difference
  $\delta^1_1-\delta^0_0$ in Generalized Chiral
  Perturbation Theory.}
\end{figure}
The three bands show the result obtained in Generalized ChPT for
$a_0^0=0.18,0.22, 0.26$, respectively. The width of the bands corresponds to
the uncertainty in the prediction. A fit of the data that exploits the
correlation between $a_0^0$ and $a_0^2$  
yields 
\bdm a_0^0=0.216\pm 0.013\, (\mbox{stat})\pm
0.004\,(\mbox{syst})\pm0.005 \,(\mbox{th})\; \citebk{Brookhaven}\co\edm
where the third error bar accounts for the theoretical uncertainties. The
result thus beautifully confirms the prediction of ChPT, 
$a_0^0=0.220\pm 0.005$. The agreement implies that more than 94\% of the pion
mass originate in the quark condensate, thus confirming that the
Gell-Mann-Oakes-Renner relation is approximately valid \citebk{PRLCGL}. 
May Generalized ChPT rest in peace. 

\section{Comparison of electromagnetic and weak form factors}
In the theoretical limit $m_u=m_d$ and in the absence of the electromagnetic
interaction, the vector current relevant for strangeness
conserving semileptonic transitions is conserved. The matrix element of this
current that shows up in the decay $\tau\rightarrow \pi^-\pi^0\, \nu_\tau$
is then determined by the electromagnetic form factor of the pion. 
In reality, however, $m_u$ differs from $m_d$ and
the radiative corrections in $\tau$ decay are different from those relevant for
$e^+e^-\rightarrow \pi^+\pi^-$. 
For the anomalous moment of the muon, $\tau$ decays are of interest
only to the extent that these isospin breaking effects are understood, so that
the e.m.~form factor can be calculated from the weak transition matrix element.

The leading isospin breaking effects
are indeed well understood: those enhanced by the small energy denominator
associated 
with $\omega$ exchange, which are described by the factor $G_3(s)$ introduced
in section \ref{pion form factor}. 
As these do not show up in
the weak transition matrix element, they must be corrected for when
calculating the 
electromagnetic form factor from $\tau$ decays.

There is another effect that shows up in the process $e^+e^-\rightarrow
\pi^+\pi^-$, but does not affect $\tau$ decay: vacuum polarization in the
photon propagator, as illustrated by the graphs below:
\begin{figure}[h]
\hspace{-1em}{\includegraphics[width=3.3cm]{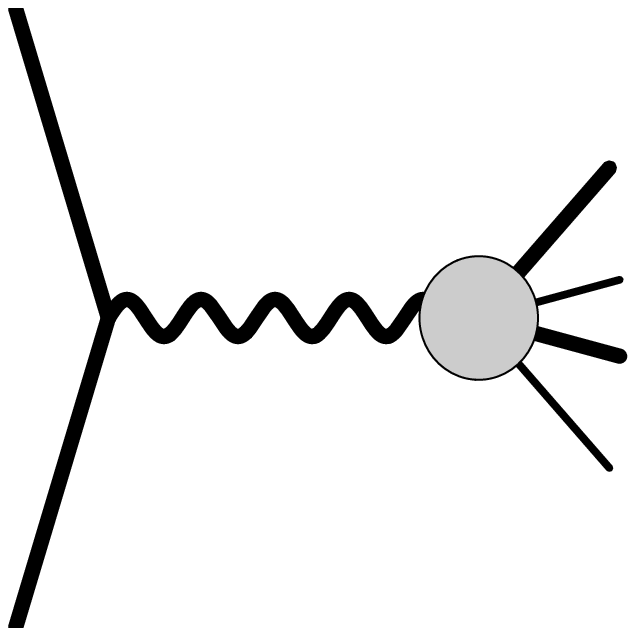}}
\raisebox{2.8em}{$+$}\hspace{-2em}\raisebox{0.25em}
{\includegraphics[width=4cm]{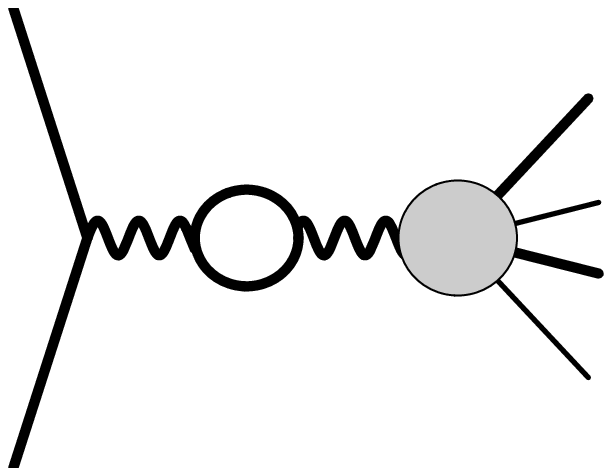}}\raisebox{2.8em}{$+$}\hspace{-2em}
\raisebox{0.25em}{\includegraphics[width=4cm]{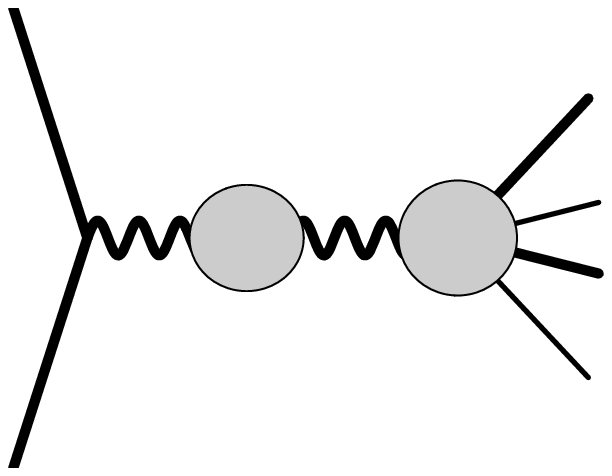}}
\end{figure}

\noindent The same graphs also show up in the magnetic moment of the muon:
\begin{figure}[thb]
\rule{1em}{0em}
\includegraphics[width=2.2cm]{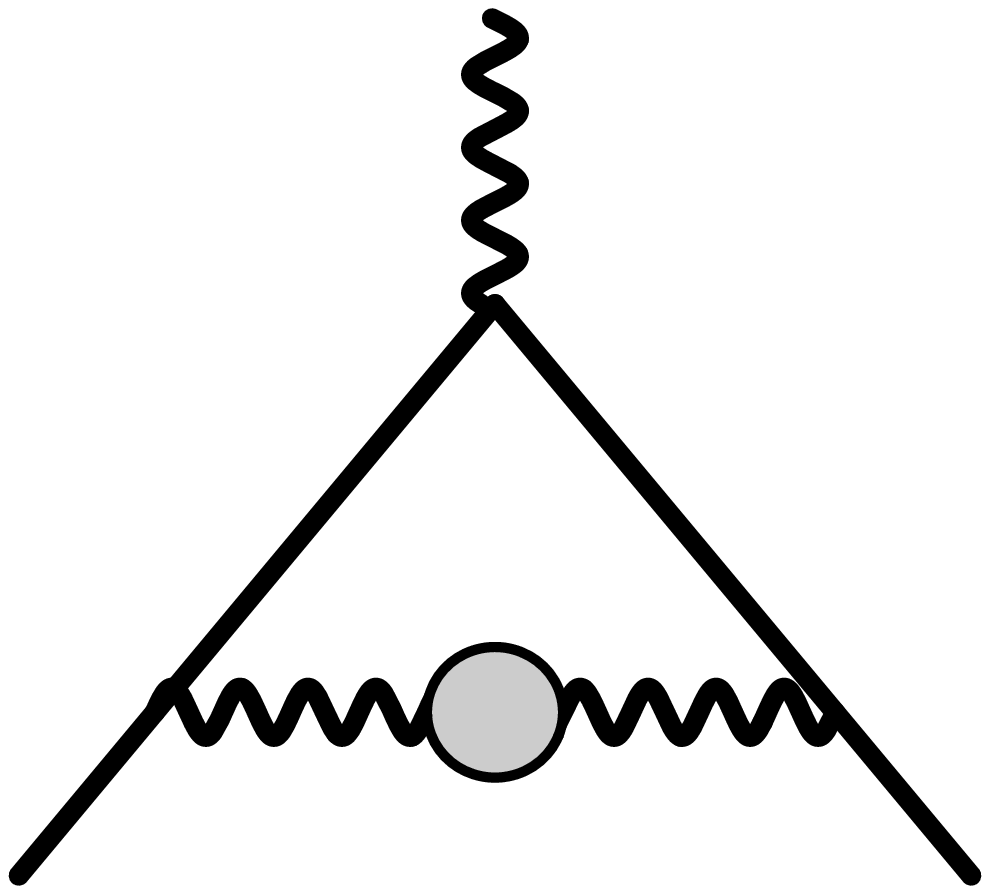}\hspace{0.5em}\raisebox{1.5em}{$+$}
\hspace{0.5em}
\includegraphics[width=2.2cm]{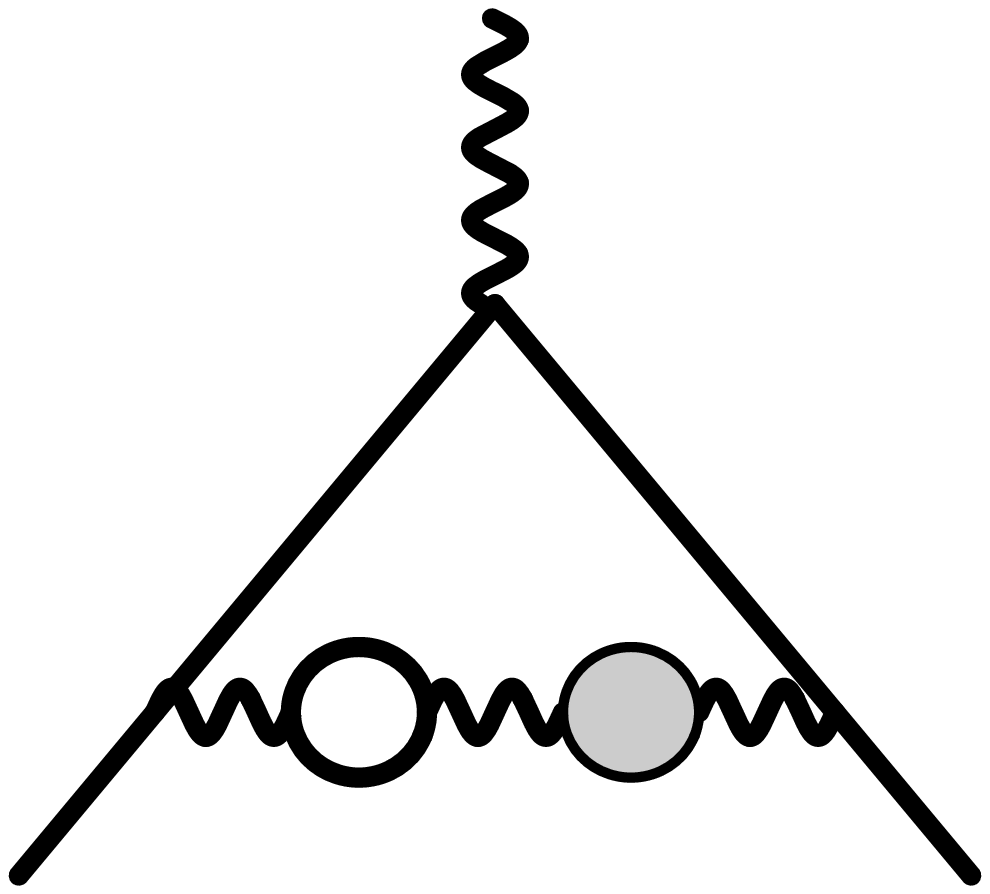}\hspace{0.5em}\raisebox{1.5em}{$+$}
\hspace{0.5em}
\includegraphics[width=2.2cm]{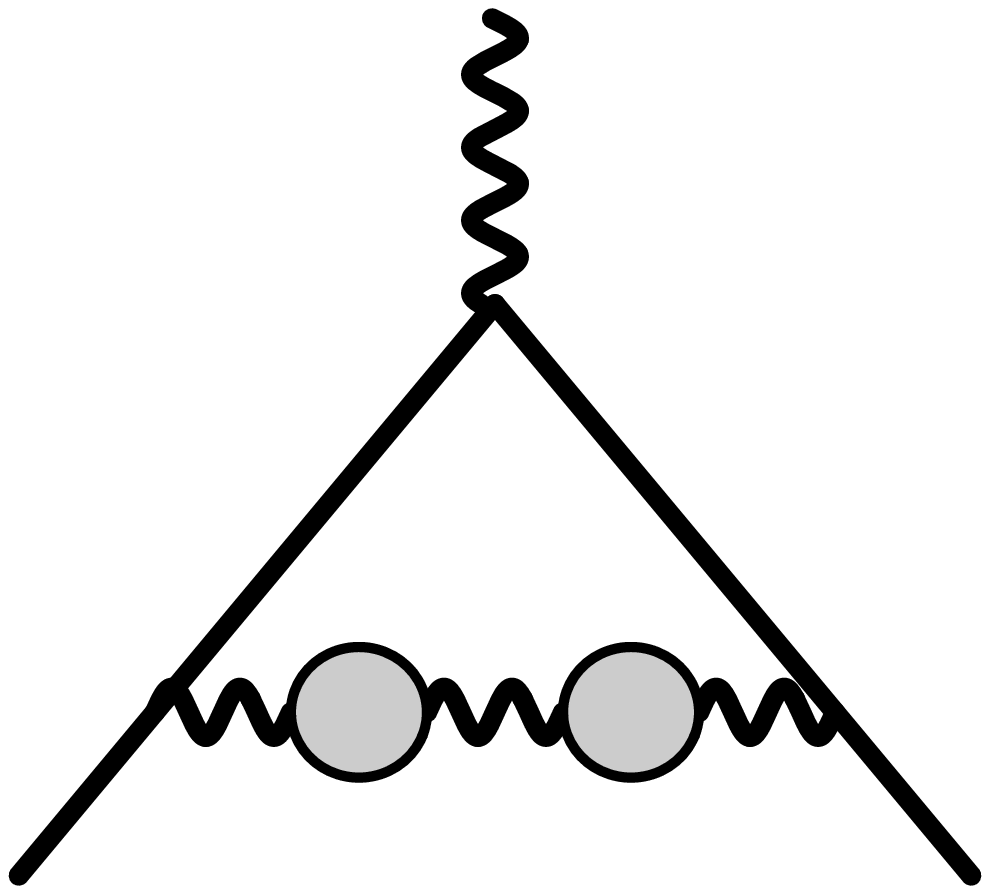}\raisebox{1.5em}{$+\;\ldots$}
\end{figure}

\noindent To avoid double counting, the data on the reaction
$e^+e^-\rightarrow\pi^+\pi^-$ must be corrected for vacuum polarization,
multiplying the cross section by the factor $|1+\Pi(s)|^2$. In the timelike
region, this factor is less than 1, so that the correction reduces
the magnitude of the form factor. In the $\omega$ region, the vacuum
polarization due to $3\pi$ 
intermediate states generates a pronounced structure in $\Pi(s)$. 
While below that energy, the correction is of order 1\%, it reaches
about 7\% immediately above the $\omega$ and then decreases to about 3\%
towards the upper end of the range covered by the CMD2 data \citebk{CMD2}.

Unfortunately, applying the two corrections just discussed,
the results for the form factor 
obtained with $e^+e^-$ collisions are systematically lower than those
found in $\tau$ decays [\citeno{ALEPH}--\citeno{OPAL}]. 
The two phenomena mentioned above are not the only isospin breaking
effects. Radiative corrections must be applied, to $e^+e^-$ collisions
\citebk{Hoefer Gluza Jegerlehner} as well as
to $\tau$ decays \citebk{Cirigliano Ecker Neufeld} and terms of order
$m_u-m_d$  
need to be estimated as well. I do not know of a mechanism, however,
that could give rise to an additional isospin breaking effect of the required
order of magnitude.  

One way to quantify the discrepancy is to assume that the uncertainties in the
overall normalization 
of some of the data are underestimated. Indeed, if the normalization of either
the rate of the decay $\tau\rightarrow \pi^-\pi^0\,\nu_\tau$ or the cross section
of the reaction $e^+e^-\rightarrow \pi^+\pi^-$ are treated as free
parameters, the problem 
disappears. The renormalization, however, either lowers the ALEPH and
CLEO data by about 4\% or lifts the CMD2 data by this amount.

While completing this manuscript, a comparison
of the $\tau$ and $e^+e^-$ data appeared \citebk{DEHZ}, where
the problem is discussed in detail. One way to put the discrepancy in evidence
is to compare the observed 
rate of the decay $\tau\rightarrow \pi^-\pi^0\nu_\tau$
with the prediction that follows from the data on the reaction
$e^+e^-\rightarrow \pi^+\pi^-$ 
if the known sources of isospin breaking are
accounted for. Using the observed lifetime of the $\tau$, the prediction for
the branching ratio of the channel $\tau\rightarrow\pi^-\pi^0\,\nu_\tau$ reads
$B_{\mbox{\tiny pred}}=0.2408\pm 0.0031$. The observed value, 
$B_{\mbox{\tiny obs}}=0.2546\pm 0.0012$, differs from this number
at the $4\,\sigma$ level \citebk{DEHZ}. 

A difference
between the $\tau$ and $e^+e^-$ results existed before, but with the new CMD2
data, where the hadronic part of vacuum polarization is now corrected
for, the disagreement has become very serious. 
The problem also manifests itself in the values for the $\rho$ width 
quoted by the
Particle Data Group \citebk{PDG}: the value obtained by CMD2 
is substantially lower than the results found by the 
ALEPH and CLEO collaborations.  

So, unless the Standard Model fails here,
either the experimental results for the electromagnetic form factor or those
for the weak form factor must be incorrect. The preliminary data from KLOE
appear to confirm the CMD2 results, but the uncertainties to be attached to
that determination of the electromagnetic form factor yet remain to be
analyzed. 

Currently, the discrepancy between the $e^+e^-$ and $\tau$ data prevents a
test of the Standard Model prediction for the magnetic moment of
the muon at an accuracy that would be comparable to the experimental value.
The result for the contribution to the muon anomaly due to
hadronic vacuum polarization depends on whether the
$e^+e^-$ data or the $\tau$ data are used to evaluate the electromagnetic form
factor -- according to ref.~\citebk{DEHZ}, the corresponding central values
differ by $17.2\times 10^{-10}$. Compared to this, the estimate 
$(8\pm 4)\times 10^{-10}$ for the contribution from
hadronic light-by-light scattering [\citeno{Knecht 
  Nyffeler}--\citeno{Rafael}] is a rather precise 
number.\footnote{The physics --
in particular the sign -- of the light-by-light contribution is well
understood: The low energy 
expansion of this term is dominated by a logarithmic singularity with known
residue 
\citebkcap{Knecht Nyffeler Perrottet de Rafael,Ramsey-Musolf Wise}. In my
opinion, 
the quoted error estimate is conservative.}

\section{Asymptotic behaviour}
The behaviour of the form factor for large spacelike momenta can be predicted
on the basis of perturbative QCD \citebk{asymptotics of form factor}:
\bea\label{pQCD} F(-Q^2)&=&\frac{64 \pi^2 F_\pi^2}{\beta_0 \,Q^2
  L_Q}\,\left\{1+B_2\,L_Q^{-\mbox{\tiny $\frac{50}{81}$}}+
B_4\,L_Q^{-\mbox{\tiny $\frac{364}{405}$}}+O(L_Q^{-1})
\right\}^2\co\\
L_Q&=& \ln \frac{Q^2}{\Lambda^2}\co\hspace{2em}
\beta_0=11-\frac{2}{3}N_f\fs\nonumber\eea
The leading asymptotic term only involves the pion decay constant. The
coefficients 
$B_2, B_4\ldots$ of the fractional logarithmic corrections are 
related to the pion distribution amplitude or null plane wave function
$\psi(x,\mu)$, which is a function of the momentum fraction $x$ of the quark 
and depends on the scale $\mu$. 
Normalizing the wave function to the
pion decay constant, 
the expansion in terms of Gegenbauer polynomials starts with
\bea \psi(x,\mu)&=& 6\, x\,(1-x)\left
\{1+B_2\,L_\mu^{-\mbox{\tiny $\frac{50}{81}$}}
\, C_2^{\frac{3}{2}}(2x\!-\!1) +B_4\,L_\mu^{-\mbox{\tiny $\frac{364}{405}$}}
\, C_4^{\frac{3}{2}}(2x\!-\!1)+\ldots\right\}\no
C_2^{\frac{3}{2}}(z)&=&\mbox{$\frac{3}{2}$}\,(5z^2-1)\co\hspace{1em}
C_4^{\frac{3}{2}}(z)=\mbox{$\frac{15}{8}$}\,(21 z^4-14
z^2+1)\co\hspace{1em}\ldots
\nonumber\eea
The wave function cannot be calculated within perturbative QCD and the 
phenomenological information about the size of the coefficients 
$B_2,B_4,\ldots$ is meagre. It is therefore of interest to see whether the
data on the form factor allow us to estimate these terms.

In the representation (\ref{three factors}), the asymptotic behaviour of the
form factor can be accounted for as follows. One first continues
the asymptotic formula (\ref{pQCD}) into the timelike region and reads off
the asymptotic behaviour of the phase of the form factor:
\bdm \phi(s)\rightarrow \pi+ \frac{\pi}{\ln \frac{s}{\Lambda^2}} 
+ \ldots\edm 
If the  asymptotic behaviour of the phase used for the Omn\`es factor agrees
with this, then the Omn\`es formula (\ref{G2})
ensures that the ratio $F(s)/G_2(s)$ approaches a constant for large spacelike
momenta. The value of the constant is determined by $F_\pi$ and by the
behaviour of the phase shift at nonasymptotic energies. This implies that the 
background term $G_4(s)$ tends to a known constant for large values
of $s$, or equivalently, for $z\rightarrow -1$. 

The corrections involving fractional
logarithmic powers can also be accounted for with a suitable contribution to
the phase. For the asymptotic expansion not to contain 
a term of order $s^{-1/2}$, the derivative of
$G_4(s)$ with respect to $z$ must vanish at $z=-1$. This then yields a
representation of the form factor for which the asymptotic behaviour agrees
with perturbative QCD, for any value of the coeffcients $B_2$ and $B_4$.

We have analyzed the experimental information with a representation of this
type,  including data in the spacelike region, as well as those
available at large timelike momenta [\citeno{Volmer}-\citeno{Psi}].
The numerical analysis yet needs to be completed and compared with the
results in the literature
(for a recent review and references, see for instance
\citebk{Bijnens Khodjamirian}).  
Our preliminary results are:
If the fractional logarithmic powers are dropped ($B_2=B_4=0$), we find
that 
the asymptotic formula is reached only at academically high energies.
With the value for $B_2$ proposed by Chernyak and Zhitnitsky, the situation
improves. For the asymptotic behaviour to set in early, an even larger value
of $B_2$ appears to be required. 

This indicates that the leading asymptotic term can dominate the behaviour
only for very high energies.  A direct 
comparison of that term with the existing data, which only cover small values
of $s$ does therefore not appear to be meaningful.

\section{Zeros and sum rules}

Analyticity subjects the form factor to strong constraints. Concerning the
asymptotic behaviour, I assume that $|\ln F(s)|$ at most grows
logarithmically for $|s|\rightarrow \infty$, in any direction of the complex
$s$-plane. This amounts to the requirement
that a) for a sufficiently large value of $n$, the quantity $|F(s)/s^n|$
remains bounded and b) the phase of $F(s)$ at most grows logarithmically, so
that the real and imaginary parts of $F(s)$ do not oscillate too rapidly at
high energies. In view of asymptotic freedom, I take these properties for
granted. If the form factor does not have zeros, the function 
\bdm \psi(s)= \frac{1}{(s_2-s)^{\frac{3}{2}}}\ln \frac{F(s)}{F(s_2)}\edm
is then analytic in the cut plane and tends to zero for $|s|\rightarrow\infty$.
The branch point singularity at threshold is of the type 
$\psi(s)\sim (s-s_2)^{-\frac{1}{2}}$. Hence $\psi(s)$ obeys the unsubtracted
dispersion relation
\bdm \psi(s)=\frac{s}{\pi}\int_{s_2}^\infty\!dx\,\frac{\mbox{Im}\,\psi(x)}
{x\,(x-s)}\edm
in the entire cut plane. The discontinuity across the cut is determined by the
magnitude of the form factor:
\bdm \mbox{Im}\,\psi(s)= -\frac{1}{(s-s_2)^{\frac{3}{2}}}\,
\ln\,
\rule[-0.8em]{0.04em}{2.2em}\,\frac{F(s)} 
{F(s_2)}\,\rule[-0.8em]{0.04em}{2.2em} 
\hspace{2em}s>s_2\fs\edm
Hence the above dispersion relation amounts to a representation of the form
factor in terms of its magnitude in the timelike region:
\bea \label{dispF}\hspace{-2em}F(s)= \rule[-0.2em]{0.04em}{1em}\,
F(\s)\, \rule[-0.2em]{0.04em}{1em}\,\exp\left\{
-\frac{(\s-s)^{\frac{3}{2}}}{\pi}\int_{\s}^\infty \!\!
\frac{dx}{(x-\s)^{\frac{3}{2}}(x-s)}\,\ln\,
\rule[-0.8em]{0.04em}{2.2em}\,\frac{F(x)} 
{F(\s)}\,\rule[-0.8em]{0.04em}{2.2em}\,\right\}.\eea
The relation implies, for instance, that the 
magnitude of the form factor in the timelike region also determines the
charge radius.\footnote{For a detailed discussion of the interrelation between
  the behaviour in the spacelike and timelike regions, in particular also in
  the presence of zeros, I refer to \citebkcap{Geshkenbein}.}

Since the value at the origin is the charge,
$F(0)=1$, the magnitude of the form factor must obey the sum rule
\bea \label{SR1} 
\frac{8 M_\pi^3}{\pi}\int_{\s}^\infty
\frac{ds}{s(s-\s)^\frac{3}{2}}\;\ln \,
\rule[-0.8em]{0.04em}{2.2em}\,\frac{F(s)} 
{F(\s)}\,\rule[-0.8em]{0.04em}{2.2em}=\ln\,|F(\s)|\fs\eea
A second sum rule follows from the asymptotic properties. For the quantity
$\ln F(s)$ not to grow more rapidly than the logarithm of $s$,
the function $\psi(s)$ must tend to zero more rapidly than $1/s$. Hence
the magnitude of the form factor must obey the condition 
\bea\label{SR2} \frac{2 M_\pi}{\pi}\int_{\s}^\infty 
\frac{ds}{(s-\s)^{\frac{3}{2}}}\,\ln \,
\rule[-0.8em]{0.04em}{2.2em}\,\frac{F(s)} 
{F(\s)}\,\rule[-0.8em]{0.04em}{2.2em}=0\fs\eea
The relations
(\ref{SR1}) and (\ref{SR2}) are necessary and sufficient for the existence of
an analytic continuation of the boundary values of $|F(s)|$ on the cut
that (a) is free
of zeros, (b) satisfies the condition $F(0)=1$ and (c) behaves properly
for large values of $|s|$. 

The above relations only hold if the form factor does not have zeros.  
In the scattering amplitude, zeros necessarily occur, as a 
consequence of chiral symmetry -- indeed, the main low energy properties of the
scattering amplitude may be viewed as consequences of the 
Adler zeros \citebk{Pennington Portoles}. 
For the form factor, however, chiral perturbation theory implies that zeros
can only occur outside the range where the low energy expansion holds: For the
form factor to vanish, the higher order contributions must cancel the leading
term of the chiral perturbation series.

In quantum mechanics, the form factor represents the Fourier transform of the
charge density. For the
ground state of the hydrogen atom, for instance, 
the charge density
of the electron cloud is proportional to the square of the wave function,
which does 
decrease with distance, so that the corresponding form factor
is positive in the spacelike region. It does not have any complex zeros,
either. The wave functions of radially excited states, on the other hand,
contain nodes, so that the form factor does exhibit zeros. 
Qualitatively, I expect
the properties of the pion charge distribution to be similar to the one of the
electron in the ground state of the hydrogen atom --
in the null plane picture, the form factor again represents the Fourier 
transform of the square of the wave function \citebk{null plane}. 
In simple models such as those described in \citebk{Jaus}, 
the form factor is free of zeros.

The hypothesis that the form factor does not contain 
zeros can be tested experimentally: The sum rules (\ref{SR1})
and (\ref{SR2}) can be evaluated with the data on the magnitude
of the form factor. The evaluation confirms that the sum rules do hold within
the experimental errors, but in the case of the slowly convergent sum rule
(\ref{SR2}), these are rather large. Alternatively, we may examine the
properties of the 
form factor obtained by fitting the data with the representation (\ref{three
  factors}). By construction, the first two factors in that representation are
free from zeros, but the term $G_4(s)$ may or may not have zeros.
In fact, as we are representing this term by a polynomial in
the conformal variable $z$, it necessarily contains zeros in the $z$-plane --
their number is determined by the degree of the polynomial. The question is
whether some of these occur on the physical sheet of the form factor, 
that is on the unit disk $|z|\leq 1$. The answer is negative: we invariably
find that all of the 
zeros are located outside the disk. It is clear that zeros at large values of
$|s|$ cannot be ruled out on the basis of experiment.
In view of asymptotic freedom, however, I think that such zeros are excluded
as well. 

\section{Conclusion}
The recent progress in our understanding of $\pi\pi$ scattering provides a
solid basis for the low energy analysis of the pion form factor.
The main problem encountered in this framework is an
experimental one: the data on the processes $e^+e^-\rightarrow \pi^+\pi^-$ and
$\tau\rightarrow \pi^-\pi^0\,\nu_\tau$ are not consistent with our
understanding of isospin breaking. 
If the data are correct, then this represents a very 
significant failure of the Standard Model -- or at least of our understanding
thereof. The discrepancy must be clarified also 
in order for the accuracy of the Standard Model prediction to become
comparable with the fabulous precision at which 
the magnetic moment of the muon has been measured.

\section*{Acknowledgment}It is a great pleasure to thank Volodya Eletsky,
Misha Shifman and  
Arkady Vainshtein for
their warm hospitality. I very much
profited from the collaboration with Irinel
Caprini, Gilberto Colangelo, Simon Eidelman, J\"urg Gasser and 
Fred Jegerlehner and
I am indebted to Andreas H\"ocker, Achim Stahl, Alexander Khodjamirian and
G\'erard Wanders for useful comments. Part of
the work reported here was carried out during the Workshop on 
Lattice QCD and Hadron Phenomenology, held in Seattle. I thank the Institute of
Nuclear Theory, University of Washington and the Humboldt Foundation
for support.


\begin{thebibliography}{99}
\bibitem{experiment}
G.~W.~Bennett {\it et al.} [Muon g-2 Collaboration],
hep-ex/0208001.

\bibitem{Schwinger}J.~Schwinger, Phys.~Rev.~{\bf 73} (1948) 416.

\bibitem{Calmet}
J.~Calmet, S.~Narison, M.~Perrottet and E.~de Rafael,\\ 
Phys. Lett. B {\bf 61} (1976) 283. 

\bibitem{Lautrup}
B.~Lautrup,
Phys.\ Lett.\ B {\bf 69} (1977) 109.

\bibitem{GL form factor}J.~Gasser and H.~Leutwyler, Nucl.~Phys.~B {\bf 250}
  (1985) 517.

\bibitem{two loops}
G.~Colangelo, M.~Finkemeier and R.~Urech,
Phys.\ Rev.\ D {\bf 54} (1996) 4403;
\\
J.~Bijnens, G.~Colangelo and P.~Talavera,
JHEP {\bf 9805} (1998) 014;
\\
P.~Post and K.~Schilcher,
hep-ph/0112352;
\\
J.~Bijnens and P.~Talavera,
JHEP {\bf 0203} (2002) 046.

\bibitem{dispersive methods}
J.~F.~Donoghue, J.~Gasser and H.~Leutwyler,
Nucl.\ Phys.\ B {\bf 343} (1990) 341;
\\
J.~Gasser and U.~Mei\ss ner, Nucl.~Phys.~B {\bf
    357} (1991) 90;
\\
J.~F.~Donoghue and E.~S.~Na,
Phys.\ Rev.\ D {\bf 56} (1997) 7073;
\\
I.~Caprini,
Eur.\ Phys.\ J.\ C {\bf 13} (2000) 471;
\\
A.~Pich and J.~Portol\'es,
Phys.\ Rev.\ D {\bf 63} (2001) 093005;
\\
J.~A.~Oller, E.~Oset and J.~E.~Palomar,
Phys.\ Rev.\ D {\bf 63} (2001) 114009.

\bibitem{Troconiz Yndurain}
J.~F.~De Troc\'oniz and F.~J.~Yndurain,
Phys.\ Rev.\ D {\bf 65} (2002) 093001.

\bibitem{Heyn Lang}M.F.~Heyn and C.B.~Lang, Z.~Phys. C {\bf 7} (1981) 169.

\bibitem{ACGL}
B.~Ananthanarayan, G.~Colangelo, J.~Gasser and H.~Leutwyler,
\\
Phys.\ Rept.\  {\bf 353} (2001) 207.

\bibitem{CGL}
G.~Colangelo, J.~Gasser and H.~Leutwyler,
Nucl.\ Phys.\ B {\bf 603} (2001) 125.

\bibitem{Descotes}
S.~Descotes, N.~H.~Fuchs, L.~Girlanda and J.~Stern,
\\
Eur.\ Phys.\ J.\ C {\bf 24} (2002) 469.

\bibitem{Roy}
S.~M.~Roy,
Phys.\ Lett.\ B {\bf 36} (1971) 353.

\bibitem{Weinberg 1966}
S.~Weinberg,
 Phys.\ Rev.\ Lett.\  {\bf 17} (1966) 616.

\bibitem{GMOR}M. Gell-Mann, R. J. Oakes and B. Renner, Phys.\ Rev.\ {\bf
 175} (1968) 2195.

\bibitem{GL}
J.~Gasser and H.~Leutwyler,
Phys.\ Lett.\ B {\bf 125} (1983) 325;\\
Annals Phys.\  {\bf 158} (1984) 142.

\bibitem{KMSF}
M.~Knecht, B.~Moussallam, J.~Stern and N.~H.~Fuchs,
\\
Nucl.\ Phys.\ B  {\bf 457} (1995) 513.
ibid.\ B {\bf 471} (1996) 445.

\bibitem{Brookhaven}
S.~Pislak {\it et al.}  [BNL-E865 Collaboration],
Phys.\ Rev.\ Lett.\  {\bf 87} (2001) 221801.

\bibitem{PRLCGL}
G.~Colangelo, J.~Gasser and H.~Leutwyler,
Phys.\ Rev.\ Lett.\  {\bf 86} (2001) 5008.

\bibitem{CMD2}
R.~R.~Akhmetshin {\it et al.}  [CMD-2 Collaboration],
\\
Phys.\ Lett.\ B {\bf 527} (2002) 161.

\bibitem{ALEPH} R.~Barate {\it et al.} [ALEPH Collaboration], 
Z.~Phys.~C {\bf 76} (1997) 15.

\bibitem{CLEO}K.~W.~Edwards {\it et al.}  [CLEO Collaboration],
Phys.\ Rev.\ D {\bf 61} (2000) 072003.

\bibitem{OPAL}
K.~Ackerstaff {\it et al.} [OPAL Collaboration], 
Eur.\ Phys.\ J.\  C {\bf 7} (1999) 571.

\bibitem{Hoefer Gluza Jegerlehner}
A.~Hoefer, J.~Gluza and F.~Jegerlehner,
Eur.\ Phys.\ J.\ C {\bf 24} (2002) 51.

\bibitem{Cirigliano Ecker Neufeld}
V.~Cirigliano, G.~Ecker and H.~Neufeld,
Phys.\ Lett.\ B {\bf 513} (2001) 361;
\\
JHEP {\bf 0208} (2002) 002.

\bibitem{DEHZ}M.~Davier, S.~Eidelman, A.~H\"ocker and Z.~Zhang, hep-ph/0208177.

\bibitem{PDG}K. Hagiwara {\it et al.} [Particle Data Group], 
Phys. Rev. D {\bf 66} (2002) 010001. 

\bibitem{Knecht Nyffeler}M.~Knecht and A.~Nyffeler,
Phys.\ Rev.\ D {\bf 65} (2002) 073034.

\bibitem{Knecht Nyffeler Perrottet de Rafael}
M.~Knecht, A.~Nyffeler, M.~Perrottet and E.~de Rafael,
\\
Phys.\ Rev.\ Lett.\  {\bf 88} (2002) 071802.

\bibitem{Ramsey-Musolf Wise}
M.~Ramsey-Musolf and M.~B.~Wise,
Phys.\ Rev.\ Lett.\  {\bf 89} (2002) 041601.

\bibitem{Rafael}
E.~de Rafael,
hep-ph/0208251.

\bibitem{asymptotics of form factor}
G.~P.~Lepage and S.~J.~Brodsky,
 Phys.\ Lett.\ B {\bf 87} (1979) 359;\\ 
 Phys.\ Rev.\ D  {\bf 22} (1980) 2157;
\\
A.~V.~Efremov and A.~V.~Radyushkin,
Phys.\ Lett.\ B  {\bf 94} (1980) 245; 
\\
Theor.\ Math.\ Phys.\  {\bf 42} (1980) 97;
\\ 
V.~L.~Chernyak and A.~R.~Zhitnitsky,
JETP Lett.\  {\bf 25} (1977) 510; 
\\
 Sov.\ J.\ Nucl.\ Phys.\  {\bf 31} (1980) 544;\\ 
G.~R.~Farrar and D.~R.~Jackson,
Phys.\ Rev.\ Lett.\  {\bf 43} (1979) 246.

\bibitem{Volmer}
J.~Volmer {\it et al.}  [The Jefferson Lab F(pi) Collaboration],
Phys.\ Rev.\ Lett.\  {\bf 86} (2001) 1713.

\bibitem{NA7}S.~R.~Amendolia {\it et al.}, Nucl.~Phys.~B {\bf 277} (1986) 168.

\bibitem{Bebek}C.~J.~Bebek {\it et al.}, Phys.~Rev.~D {\bf 17} (1978) 1693.

\bibitem{Bollini} D.~Bollini {\it et al.}, Nuovo Cim.~Lett.~{\bf 14} (1975)
  4188. 

\bibitem{Psi}
J.~Milana, S.~Nussinov and M.~G.~Olsson,
Phys.\ Rev.\ Lett.\  {\bf 71} (1993) 2533.

\bibitem{Bijnens Khodjamirian}
J.~Bijnens and A.~Khodjamirian,
hep-ph/0206252.

\bibitem{Geshkenbein}B.~V.~Geshkenbein,
Yad.~Fis.~{\bf 9} (1969) 1932; Yad.~Fis.~{\bf 13} (1971) 1087; 
Z.~Phys.~C {\bf 45} (1989) 351;
Phys.\ Rev.\ D {\bf 61} (2000) 033009.

\bibitem{Pennington Portoles} M.R.\ Pennington and J.\ Portol\'es,
Phys.\ Lett.\ B {\bf 344} (1995) 399.

\bibitem{null plane}H.~Leutwyler, Nucl.\ Phys.\ B {\bf 76} (1974) 413.

\bibitem{Jaus}W.~Jaus, Phys.\ Rev.\ D {\bf 44} (1991) 2851.


\end{thebibliography}
\end{document}